\documentclass[12pt]{article}
\usepackage{graphicx}
\def\be{\begin{equation}}
\def\ee{\end{equation}}
\def\bea{\begin{eqnarray}}
\def\eea{\end{eqnarray}}

\begin{document}
\begin{titlepage}
\begin{center}
{\Large \bf Theoretical Physics Institute, University of Alberta \\}
{\Large \bf and \\}
{\Large \bf William I. Fine Theoretical Physics Institute \\
University of Minnesota \\}
\end{center}
\vspace{0.2in}
\begin{flushright}
Alberta Thy 25-17\\
FTPI-MINN-17/14 \\
UMN-TH-3633/17 \\
August 2017 \\
\end{flushright}
\vspace{0.3in}

\begin{center}
{\LARGE \bf Stability of tetrons \\}
\vspace{0.2in}
{\bf  Andrzej Czarnecki$^{1}$, Bo Leng$^{1}$ and M.B. Voloshin$^{2,3,4}$  \\ }
$^1$Department of Physics, University of Alberta, Edmonton, Alberta, Canada T6G 2E1\\
$^2$William I. Fine Theoretical Physics Institute, University of 
Minnesota,\\ Minneapolis, MN 55455, USA \\
$^3$School of Physics and Astronomy, University of Minnesota, Minneapolis, MN 55455, USA \\ 
$^4$Institute of Theoretical and Experimental Physics, 117218 Moscow,  Russia 
\\[0.2in]
\end{center}

\vspace{0.2in}

\begin{abstract}
We consider the interactions in a mesonic system, referred here to as `tetron', consisting of two heavy quarks and two lighter antiquarks (which may still be heavy in the scale of QCD), i.e.~generally $Q_a Q_b \bar q_c \bar q_d$, and study the existence of bound states below the threshold for decay into heavy meson pairs.  
At a small ratio of the lighter to heavier quark masses an expansion parameter arises for treatment of the binding in such systems. We find that in the limit where all the quarks and antiquarks are so heavy that a Coulomb-like approximation can be applied to the gluon exchange between all of them, such bound states arise when this parameter is below a certain critical value. We find the parametric dependence of the critical mass ratio on the number of colors $N_c$, and confirm this dependence by numerical calculations. In particular there are no stable tetrons when all constituents have the same mass. 
We discuss an application of a similar expansion in the large $N_c$ limit to realistic systems where the antiquarks are light and their interactions are nonperturbative. In this case our findings are in agreement with the recent claims from a phenomenological analysis that a stable $b b \bar u \bar d$ tetron is likely to exist, unlike those where one or both bottom quarks are replaced by the charmed quark. 
\end{abstract}
\end{titlepage}

\section{Introduction}
Multiquark hadrons, whose internal structure apparently goes beyond the standard template of three quark baryons and quark-antiquark mesons, have recently been observed in various experiments (for a recent review see e.g.~\cite{als,ghmwzz}). All such exotic hadrons found so far contain a heavy $b$ or $c$ quark and a corresponding antiquark. For this reason they all are unstable with respect to annihilation of a heavy quark-antiquarks pair, even though their rate of dissociation into conventional hadrons can be small. There is a whole spectrum of theoretical models for description of such resonances. In particular the most discussed models for the mesonic ones are the molecular~\cite{vo,ghmwzz}, the tetraquark (a recent review can be found in Ref.~\cite{epp}) and the `hadro-quarkonium'~\cite{mv07,dv}. A different kind of phenomenology of multiquark hadrons would be accessible if there existed systems made of two heavy quarks (as opposed to a quark-antiquark pair) and two lighter antiquarks:
$Q_a Q_b \bar q_c \bar q_d$, that would be bound below the threshold for dissociation into a pair of $Q \bar q$ mesons. Such hadrons have been discussed within the quark model for quite some time~\cite{art,hl}, and the lightest of them can decay only through the weak interaction. In view of special properties of such systems we call them here `tetrons' implying that they in fact are `stable' mesons made of four constituents. 

A recent revival of the interest in tetrons is inspired by the observation~\cite{lhcb}
by LHCb of a doubly charmed baryon $\Xi_{cc}^{++} \sim ccu$. The measurement of the mass of the baryon at about 3621\,MeV has provided an estimate of the effective mass of the heavy quark pair $cc$ (with the interaction between the quarks in the color antitriplet state), and thus  an input into phenomenological models~\cite{kr,eq}. 

The latter models are based on the picture~\cite{hl}, where due to the attraction in the color-antisymmetric state, the heavy quark pair forms a compact, in fact a point-like, bound state. This bound state then acts essentially as a heavy antiquark and binds either with a light quark to form a baryon, e.g.~$\Xi_{cc}^{++}$, or with a light antiquark pair to form a tetron, e.g.~$Q_a Q_b \bar u \bar d$. Since the latter binding is similar to that in respectively a heavy meson and a heavy (anti)baryon, by applying the known mass differences, e.g.~between $\Lambda_c$ and $D$, or between $\Lambda_b$ and the $B$-meson, the masses of possible tetrons containing $cc$, or $bb$, or $bc$ heavy quark pair can be estimated. In this way it has been argued~\cite{kr,eq} that there are no stable tetrons with $cc$  heavy quark pair, but there definitely is a $bb \bar u \bar d$ one,  well below the $B^- \bar B^0$ threshold, and also likely similar weakly decaying strange tetrons~\cite{eq} $bb \bar s \bar q$ with $q$ standing for either $u$ or $d$. 
Numerical evidence for such states has been established in lattice nonrelativistic QCD \cite{Francis:2016hui}, as well as using the approximation of static $b$-quarks \cite{Bicudo:2016ooe}. 
 (The conclusion about existence of mixed bottom-charm tetrons $b c \bar q \bar q$ is not conclusive in Ref.~\cite{kr} and negative in \cite{eq}.)

It is clear however that the similarity of the interaction in a tetron to that in an (anti)baryon, where a heavy antiquark is replaced by a compact color-antisymmetric pair of heavy quarks is not exact. One simple reason for a deviation is the spin-dependent interaction, which is suppressed for heavy quarks and which to some extent can be accounted for~\cite{eq}. The other (and less tractable) reason is that the heavy quark pair has a finite size with the most important effect being a flip of the color state from antisymmetric to symmetric (with the corresponding change of the color of the light antiquark pair). The existence of these configurations was recognized in the previous studies~\cite{art,
zsgr,vvb} and was taken into account in a series of approximations.

 In what follows we treat  the mixing of the color configurations explicitly within an expansion in the ratio of the distance between the heavy quarks to the characteristic distance to the light antiquarks. The point-like limit~\cite{hl,kr,eq} is the first term in this expansion. It naturally appears that for sufficiently heavy quark pair with the (reduced) mass $M$, the characteristic size of the bound state is proportional to $1/M$, while the distance scale for the light (massless) antiquarks in a tetron is set by $\Lambda_{QCD}$, so that the ratio of the distance scales is proportional to $\Lambda_{QCD}/M$. We will argue however that the effects of the deviation from the point-like approximation are enhanced in the limit of large number $N_c$ of colors, so that the relevant parameter for this deviation is in fact
\be
\xi = N_c^6 \left ( {\Lambda_{QCD} \over M} \right ) ^4~,
\label{npxi}
\ee
which at $N_c=3$ indicates that the point-like limit is not applicable if at least one of the heavy quarks is the charmed one. On the other hand, this limit may work with reasonably small corrections of order $\xi$ for tetrons with the $bb$ quark pair.

Furthermore, it appears that a stable tetron does not exist if the parameter $\xi$ is of order one or larger. To establish this behavior, we consider in Section 2 the limit where all the quarks and the antiquarks are asymptotically heavy, so that the relevant distances for bound states are short. One can then apply the Coulomb-like limit for the gluon exchange among all constituents, with a non-relativistic Hamiltonian describing the interplay of color configurations. The two scales are introduced in this model by considering the quarks $Q$ as having mass $M$ that is larger than the mass $m$ of the antiquarks $\bar q$. The ratio $f=m/M$ is a variable parameter.~\footnote{We consider, for simplicity, the situation where the heavier quarks are the same as well as the lighter antiquarks are the same. The consideration can be  generalized to different masses in the limit of a strong mass hierarchy  by introducing appropriate reduced masses.} The bound state problem in this model is solved by a numerical variational calculation; on the other hand it is  analyzed in terms of an expansion in the size of the heavy bound $QQ$ pair. We find that an analog of the parameter (\ref{npxi}) in this solvable model is 
\be
\xi_c= N_c^6 \, f^4~.
\label{cxi}
\ee
On the other hand we find from the numerical calculation that a stable tetron in this system exists only when the ratio $f$ is smaller than a certain critical value $f_c(N_c)$,
\be
f_c \approx a/N_c^{3 \over 2}
\label{fcn}
\ee
where the coefficient $a$ is of order one, numerically $a \approx 0.77$. It is thus plausible that the condition for existence of a stable tetron is a small value of the expansion parameter [in this model $\xi_c$ in Eq.~(\ref{cxi})] describing the deviation from the point-like model for the pair of heavy constituents.

Unlike in the solvable model with Coulomb-like forces, interactions in a system containing light $u, d,$ or $s$ quarks cannot be described by a potential. However some features of a gluon exchange can be applied to such systems in the limit of large number of colors $N_c$ with the usual assumption~\cite{tHooft} that, as $N_c$ increases, the coupling $\alpha_s$ decreases, so that the product $N_c \alpha_s$ stays of order one. We discuss the parameters describing a tetron in this limit in Section 3.

Finally,  Section 4 contains general discussion and conclusions.

\section{A solvable model with superheavy quarks}

We consider a system of two heavy quarks $Q$ with mass $M$ and two lighter (but still heavy) antiquarks $\bar q$ with mass $m$ each. For the start we assume  no statistics symmetry constraints, e.g.~assuming that the quarks are not identical, even though they have the same mass. The odd numbered positions $\vec r_1$ and $\vec r_3$ refer to the quarks, while the even  ones $\vec r_2$ and $\vec r_4$ are those for the antiquarks. The gluon exchange potential between the color constituents at positions $\vec r_i$ and $\vec r_j$ is
\be
V_{ij}= T^a_{(i)} T^a_{(j)} d_{ij}
\label{vij}
\ee
with $T^a_{(i)}$ being the color generators acting on the constituent at $\vec r_i$, and $d_{ij}$ in the Coulomb limit is given by
\be
d_{ij}= {\alpha_s \over |\vec r_i - \vec r_j|}~.
\label{dij}
\ee
The condition for the system to be colorless can be satisfied with two configurations of the sub-systems described by the color combinations: 
\bea
&& \Psi= (\bar q_{(2) \, \alpha} Q_{(1)}^\alpha) ((\bar q_{(4) \, \beta} Q_{(3)}^\beta)/N_c~, \nonumber \\
&& \Phi= (\bar q_{(4) \, \alpha} Q_{(1)}^\alpha) ((\bar q_{(2) \, \beta} Q_{(3)}^\beta)/N_c~,
\label{psiphi}
\eea
where $\alpha$ and $\beta$ are color indices in the fundamental representation of the color group $SU(N_c)$. Clearly in the $\Psi$ configuration the color singlets are $(\bar q_{(2) } Q_{(1)})$ and $(\bar q_{(4) } Q_{(3)})$ while in $\Phi$ they are $(\bar q_{(4) } Q_{(1)})$ and $(\bar q_{(2) } Q_{(3)})$. The sum of pairwise one-gluon exchanges among the four constituents results in the potential that can be written in terms of $\Psi$ and $\Phi$ as
\be
V \, \left ( \begin{array}{c}
\Psi \\
\Phi \end{array} \right ) = {1 \over 2} \, \left ( \begin{array}{cc} -{N_c^2 -1 \over N_c} \, (d_{12}+d_{34}) - {1 \over N_c} \, p & p \\
q & -{N_c^2 -1 \over N_c} \, (d_{14}+d_{23}) - {1 \over N_c} \, q
 \end{array}
\right ) \left ( \begin{array}{c}
\Psi \\
\Phi \end{array} \right )~,
\label{potpsi}
\ee
where we have used  the notation
\be
p=d_{13}-d_{23}+d_{24}-d_{14}, ~~~~~q=d_{13}-d_{34}+d_{24}-d_{12}~.
\label{fg}
\ee
The potential matrix in Eq.~(\ref{potpsi}) is not symmetric, because the color states $\Psi$ and $\Phi$ in Eq.~(\ref{psiphi}) are not orthogonal,
\be
\langle \Phi|\Psi \rangle = \langle \Psi|\Phi \rangle = {1 \over N_c}~.
\ee
Orthogonal (and normalized) states can be chosen as
\be
u={ 1 \over \sqrt{ 2 (1+1/N_c)}} \, (\Psi+\Phi), ~~~~ w= { 1 \over \sqrt{ 2 (1-1/N_c)}} \, (\Psi-\Phi)~,
\label{uv}
\ee
and the one-gluon exchange potential (\ref{potpsi}) in the basis of these states reads as
\be
V  \, \left ( \begin{array}{c}
u \\
w \end{array} \right )= - {1 \over 4} \, \left ( \begin{array}{cc} {N_c^2 -1 \over N_c} \, r - {N_c-1 \over N_c} (p+q) & \sqrt{N_c^2-1} \, s \\ \sqrt{N_c^2-1} \, s & {N_c^2 -1 \over N_c} \, r + {N_c+1 \over N_c} (p+q) 
\end{array}
\right ) \, \left ( \begin{array}{c}
u \\
w \end{array} \right )
\label{potuw}
\ee
where
\be
r= d_{12}+d_{34}+d_{14}+d_{23}~,~~~~~s=d_{12}+d_{34}-d_{14}-d_{23}~.
\ee

The Hamiltonian with the potential (\ref{potuw}) clearly has a $Z_2 \times Z_2$ symmetry under switching of the positions of the quarks, $\vec r_1 \leftrightarrow \vec r_3$, and (independently) switching the positions of the antiquarks, $ \vec r_2 \leftrightarrow \vec r_4$. The symmetry of the $u$ and $w$ components is opposite; e.g.~if the $w$ component is even under swapping of quarks then the $u$ component has to be odd. This implies that the eigenstates of the Hamiltonian can be classified in terms of the symmetry of the $w$ component: $w_{++}$, $w_{--}$, $w_{+-}$ and $w_{-+}$.

Furthermore, one can readily see that the states $u$ and $w$ contain the quark (antiquark) pair of a definite color symmetry: symmetric in $u$ and antisymmetric in $w$.~\footnote{The color and coordinate symmetry properties of the components certainly become essential for identical quarks with the constraint of the Fermi-Dirac statistics.}  In particular, at $N_c=3$ the $u$ state contains a color sextet quark (anti-sextet antiquark) pair, while the state $w$ contains the antitriplet quark (triplet antiquark) pair configuration. Thus it is the latter $w$ component that is present in the phenomenological analyses of Refs.~\cite{kr,eq}. When the heavier quarks $Q$ are close to each other, the term $d_{13}$ becomes dominant, $(p+q) \approx 2 d_{13}$, and one recovers from Eq.~(\ref{potuw}) the attraction in the color antisymmetric state:
\be
V_{13} = - {N_c+1 \over 2 N_c} \, d_{13}~.
\label{v13}
\ee
This attraction  binds the $Q$ quarks into a compact Coulomb-like system  with the size and energy becoming, at large $N_c$, 
\be
r_{QQ} \sim (M \, \alpha_s)^{-1}~, ~~~~~~~E_{QQ} \sim M \, \alpha_s^2~.
\label{reh}
\ee
Clearly, at large $M$ such distance scale is small in the scale $R_q$ of the dynamics of the lighter antiquarks in the considered system, and one can consider an expansion in the ratio $r_{QQ}/R_q$. In the zeroth order of this expansion, i.e.~at vanishing $r_{QQ}$, the off-diagonal terms in Eq.~(\ref{potuw}) vanish and there is no mixing between the $w$ and $u$ components, and thus one can set $u=0$. 
Then the leading at large $N_c$ interaction for the lighter antiquarks is that with the heavier quarks. After setting $\vec r_3 = \vec r_1$ in the proportional to $N_c$ part of the diagonal term in Eq.~(\ref{potuw}) one finds the potential
\be
V_{qQ}= - {N_c \over 2} \, (d_{12}+d_{14})~,
\label{vhl}
\ee
describing an independent Coulomb-like interaction of the two lighter antiquarks with the compact $QQ$ system. Naturally, the latter interaction corresponds to spectra of two independent $Q \bar q$ Coulomb-like quarkonia, with the distance and energy scale set as
\be
R_q \sim (m \, N_c \alpha_s)^{-1}~,~~~~~E_q \sim m \, N_c^2 \alpha_s^2~.
\label{rel}
\ee
It is also clear that the ground state in both the potential (\ref{v13}) and (\ref{vhl}) is spatially symmetric, so that the overall ground state of the tetron is of the type $w_{++}$ under the $Z_2 \times Z_2$ symmetry.

Due to the binding between the heavy quarks by the potential (\ref{v13}) the resulting four-quark system is stable under decay to two quarkonium mesons. It should be noted however that this binding is only sub leading  in terms of the  large $N_c$ counting, as can be seen by comparing the expressions (\ref{v13}) and (\ref{vhl}). Thus the discussed `hierarchy' of the binding energies is only applicable if the ratio $f$ of the masses is small enough at a fixed $N_c$. In other words there is a critical value of this ratio $f_c(N_c)$ above which the described approximation fails. In order to evaluate the behavior of $f_c(N_c)$ we consider here the effects arising at a finite ratio $r_{QQ}/R_q$. We find that the main effect arises due to non-vanishing off-diagonal elements in the potential (\ref{potuw}):
\be
\sqrt{N_c^2-1} \, (d_{12}+d_{34}-d_{14}-d_{23}) \sim N_c \, \alpha_s \, r_{QQ}/R_q^2~.
\label{voffd}
\ee
This term can be considered as small as long as the energy shift that it produces in the second order is small in comparison with either of the energy scales [in Eq.~(\ref{reh}) or Eq.~(\ref{rel})]. One can readily verify that using the energy scale $E_{QQ}$ imposes a more stringent bound on $f=m/M$:
\be
(N_c \, \alpha_s \, r_{QQ}/R_q^2)^2/E_{QQ}^2 \sim N_c^6 \, (m/M)^4 \ll 1~,
\label{ub}
\ee
so that the applicability of the discussed expansion fails at $f > f_c(N_c)$ with $f_c$ given by Eq.~(\ref{fcn}). In particular the absence of a stable bound state at larger mass ratio makes highly unlikely existence of a `double bottomonium' occasionally discussed in the literature (see e.g.~Ref.~\cite{bln,bkcv,knr}).

\begin{figure}[htb]
  \centering
  \includegraphics{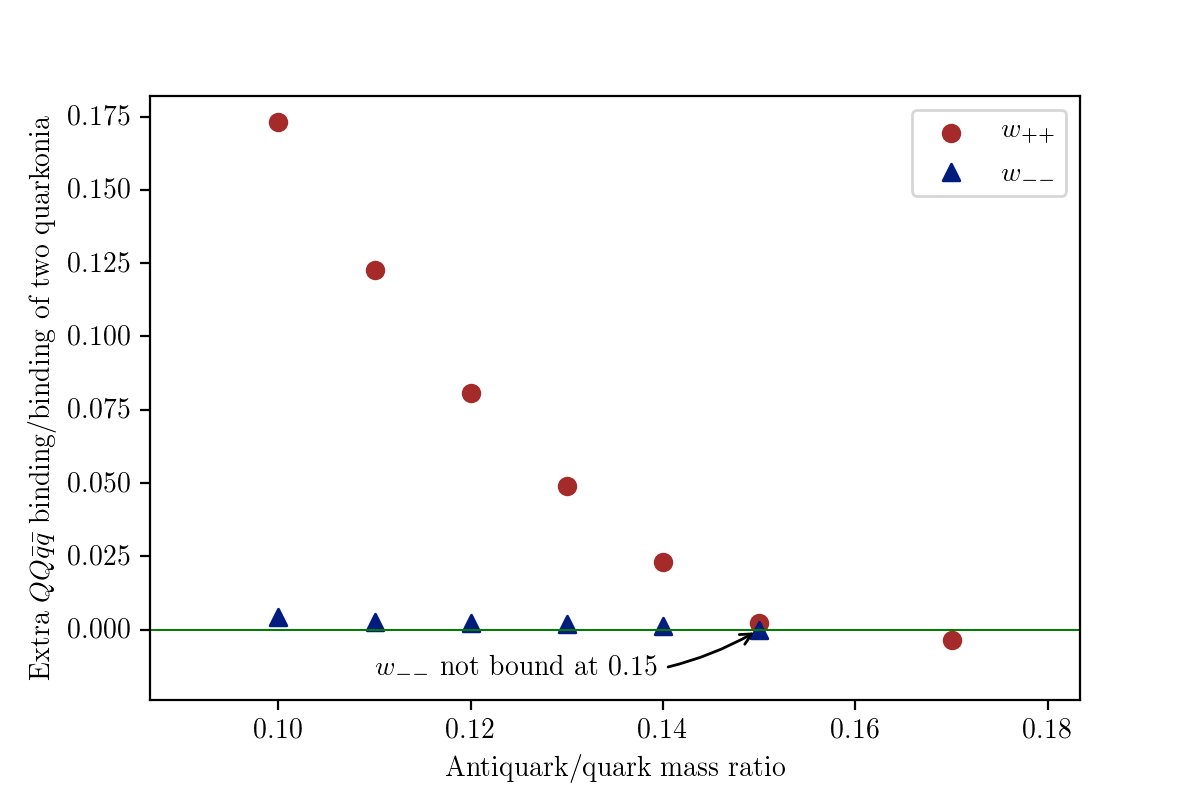}
  \caption{Extra binding energy of a tetron (in units of the total binding for two independent $Q \bar q$ mesons) as a function of the antiquark/quark mass ratio $f$. The number of colors is $N_c = 3$. The state with the symmetry $w_{++}$ (circles) is bound more strongly than $w_{--}$ (triangles). Even the state $w_{++}$ is no longer bound when the mass ratio is higher than about $f_c\simeq 0.152$. }
  \label{fig:binding}
\end{figure}

\begin{figure}[htb]
  \centering
  \includegraphics{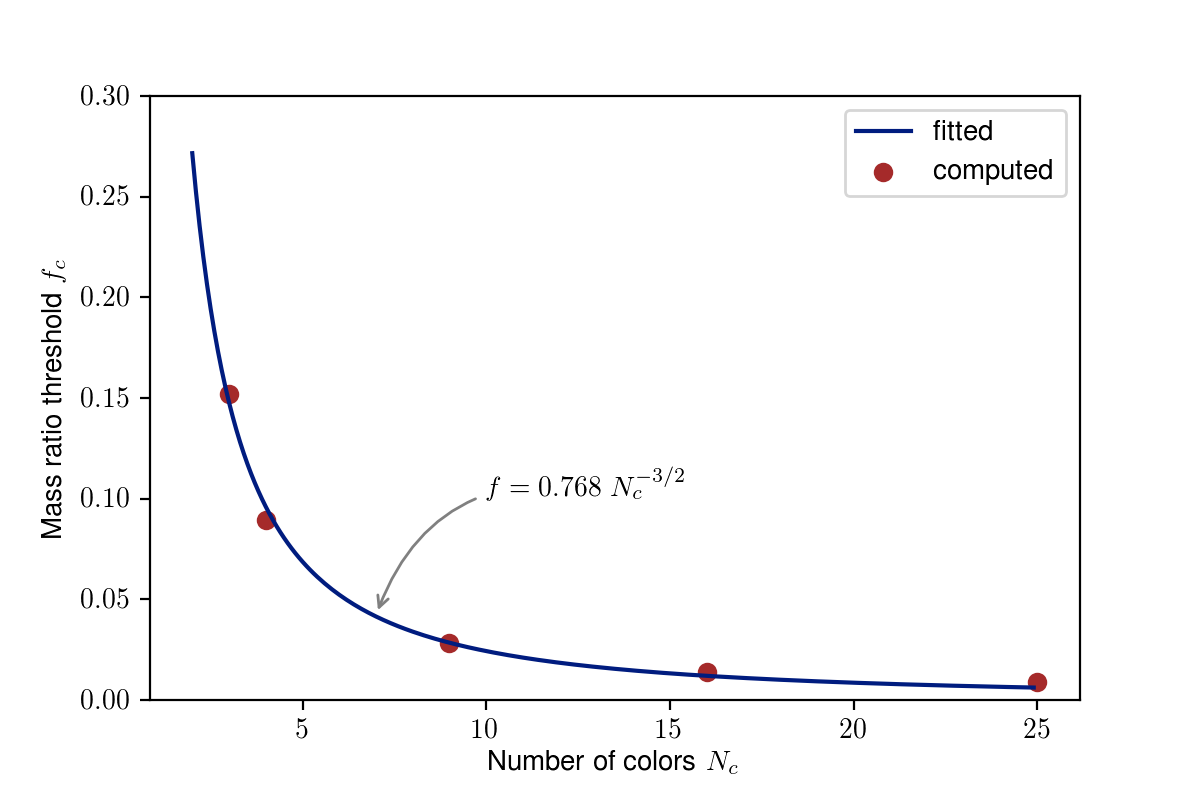}
  \caption{Mass threshold $f_c$ for tetrons as a function of the number of colors.}
  \label{fig:threshold}
\end{figure}

By performing a numerical variational calculation we find that the lowest bound state in the system is of the $w_{++}$ type and exists only when the ratio in Eq.~(\ref{ub}) is small so that the mass ratio $f$ is smaller than the critical value described by Eq.~(\ref{fcn}) (see Fig.~\ref{fig:binding}).~\footnote{We note in passing that a shallower bound state of the type $w_{--}$ also exists at sufficiently small $f$, while no bound states of mixed symmetry $w_{+-}$ or $w_{-+}$ are found in our analysis.} The results for the values of $f_c$ at which the bound state disappears at different $N_c$ are shown in 
Fig.~\ref{fig:threshold}.

We computed the data points in Fig.~\ref{fig:threshold}  with a generalization of the algorithm developed for the positronium molecule \cite{Puchalski:2008jj}. Both wave-function components $u$ and $w$ are represented as a sum of Gaussian trial functions of all six inter-particle distances. We use a basis of 200 trial functions for each of the components. Much larger bases can be employed if higher precision is warranted. A challenge in this calculation is a slow convergence very near the threshold. This explains the slight spread of the data points around the fitted curve in Fig.~\ref{fig:threshold}. 

\section{Tetron with superheavy quarks and  massless antiquarks}
A potential description, and even more in terms of a Coulomb-like potential, is not applicable for the interaction of light $u, d, s$ quarks, and other methods have to be invoked. In this section we consider a system of two very heavy quarks $QQ$ with mass $M$ each, and two massless quarks $\bar q \bar q$ (which are not necessarily identical, e.g. $\bar u \bar d$). Although literally the potential model of the previous section does not apply, some essential features of the interaction in Eq.~(\ref{potuw}) are retained, in particular a Coulomb-like potential treatment of the interaction between the heavy quarks.  Namely, the one gluon exchange between the heavy quarks still produces a compact bound state in the potential (\ref{v13}) with the relevant parameters described by Eq.~(\ref{reh}). This interaction, essential at large $M$, is however only sub-dominant in the large $N_c$ limit, in which limit the dominant effect (of order one) is the interaction between the light and heavy constituents. The heavy-light mesons $Q \bar q$ are formed and the estimate (\ref{rel}) for the relevant characteristic size   and energy scale is replaced by
\be
R_q \sim \Lambda_{QCD}^{-1}~,~~~~~~E_q \sim \Lambda_{QCD}~.
\label{renp}
\ee

Moreover, the mixing between the $w$ and $u$ components, although not describable by a potential analog of the non-diagonal components in Eq.~(\ref{potuw}), retains the following features. It is of order one in the limit of large $N_c$ and it vanishes at zero spatial separation $r_{QQ}$ between the heavy quarks. Thus one can estimate the amplitude of the mixing in the linear order of the expansion  in $r_{QQ}$ as
\be
\langle u | H | w \rangle \sim r_{QQ}/R_q^2~ \sim r_{QQ} \, \Lambda_{QCD}^2~.
\label{hoffd}
\ee
The perturbation parameter for the mixing is then evaluated as
\be
\xi \sim {|\langle u | H | w \rangle |^2 \over E_{QQ}^2} \sim {\Lambda_{QCD}^4 \over M^4 \, \alpha_s^6}~,
\label{xipar}
\ee
which results in the estimate in Eq.~(\ref{npxi}).

\section{Final remarks}

The parameter $\xi$ in Eq.~(\ref{npxi}), similarly to $\xi_c$ in (\ref{cxi}), controls the applicability of the treatment of tetron starting from a compact bound diquark made of the heavy quarks. A perturbative expansion in the spatial separation is possible when this parameter is (formally) much less than one, and generally this expansion becomes invalid once the $\xi$ is of order one. Our calculations in the solvable model with heavy quarks however revealed that not only that the expansion becomes inapplicable when $\xi_c$ is of order one, but no stable bound tetrons arise at all. We interpret this behavior as that the leading at large $N_c$ dipole force [the off diagonal terms in the potential (\ref{potuw})] results in a strong mixing between the $w$ and $u$ components. Such mixing essentially randomizes the total color of heavy diquark, so that a residual net interaction between the heavy constituents largely cancels between the color symmetric and antisymmetric configurations. We thus conclude that it is highly likely that in a more realistic tetron with light quarks the existence of a stable bound state is also controlled by the parameter $\xi$, and the stability does not exist if $\xi$ is of order one or larger.

It is certainly of a primary interest to understand the status of tetrons with the heavy constituents being the actual $b$ and $c$ quarks. Using the criterion based on the estimate in Eq.~(\ref{npxi}) one readily concludes that for the $cc \bar q \bar q$ and $b c \bar q \bar q$ systems, where the reduced mass $M$ in the heavy diquark is determined by the charm quark mass, there is essentially no chance that the parameter $\xi$ is small. Thus we confirm the finding of the earlier studies~\cite{art} that it is highly unlikely that there are stable tetrons with such quark structure. 

The parameter $\xi$ from Eq.~(\ref{npxi}) is more likely to be small enough, if $M$ is proportional to the mass of the $b$ quark. Due to the inherent uncertainty in this estimate for a nonperturbative system it would be impossible to unambiguously claim existence of stable tetrons of such type, based solely on this estimate. However we believe that there is a strong indication that if stable tetrons do exist, the only possibility for them is to be of the double bottom type. At this point we find an agreement with the conclusions based on purely phenomenological estimates in Refs.~\cite{kr,eq}. It is certainly understood~\cite{eq} that an experimental observation of double bottom tetrons can be quite challenging. However a search for them may be well worth the effort, as the tetrons possibly present a very unconventional form of hadrons that are stable with respect to strong decay.

Clearly, the smallness of the parameter $\xi$, or its analog, requires existence of two strongly separated mass scales, whose ratio can ensure that the binding effect in the color antisymmetric state due to heavy masses is not eliminated by a larger in $N_c$ destabilizing mixing between the color states. We notice absence of such hierarchy of scales for four-quark systems with only the heavy $b$ and $c$ quarks, e.g. $bb \bar c \bar c$, so that we do not expect existence of stable tetrons of such type. The same negative conclusion applies to four-quark systems with hidden heavy flavors, such as a double bottomonium $b b \bar b \bar b$, or double charmonium, $c c \bar c \bar c$, systems.

\section*{Acknowledgement} We thank Mariusz Puchalski for sharing his unpublished code for variational calculations.  We thank Connor Stephens for collaboration at an early stage of the project. The research of A.C. and B.L. was supported by the Natural Sciences and Engineering Research Council (NSERC) of Canada, and by the Munich Institute for Astro- and Particle Physics (MIAPP) of the DFG cluster of excellence {\em Origin and Structure of the Universe}. B.L. is supported by the Canadian Institute for Nuclear Physics Undergraduate Research Scholarship. The work of M.B.V. is supported in part by U.S. Department of Energy Grant No.\ DE-SC0011842.


\begin{thebibliography}{99}

\bibitem{als} 
  A.~Ali, J.~S.~Lange and S.~Stone,
  arXiv:1706.00610 [hep-ph].
	
\bibitem{ghmwzz} 
  F.~K.~Guo, C.~Hanhart, U.~G.~Mei{\ss}ner, Q.~Wang, Q.~Zhao and B.~S.~Zou,
  arXiv:1705.00141 [hep-ph].
	
\bibitem{vo} 
  M.~B.~Voloshin and L.~B.~Okun,
  JETP Lett.\  {\bf 23}, 333 (1976)
  [Pisma Zh.\ Eksp.\ Teor.\ Fiz.\  {\bf 23}, 369 (1976)].
	
\bibitem{epp} 
  A.~Esposito, A.~Pilloni and A.~D.~Polosa,
  Phys.\ Rept.\  {\bf 668}, 1 (2016)
  doi:10.1016/j.physrep.2016.11.002
  [arXiv:1611.07920 [hep-ph]].
	
\bibitem{mv07} 
  M.~B.~Voloshin,
  Prog.\ Part.\ Nucl.\ Phys.\  {\bf 61}, 455 (2008)
  doi:10.1016/j.ppnp.2008.02.001
  [arXiv:0711.4556 [hep-ph]].
	
\bibitem{dv} 
  S.~Dubynskiy and M.~B.~Voloshin,
  Phys.\ Lett.\ B {\bf 666}, 344 (2008)
  doi:10.1016/j.physletb.2008.07.086
  [arXiv:0803.2224 [hep-ph]].
	
\bibitem{art} 
  J.~P.~Ader, J.~M.~Richard and P.~Taxil,
  Phys.\ Rev.\ D {\bf 25}, 2370 (1982).
  doi:10.1103/PhysRevD.25.2370.
	
\bibitem{hl} 
  H.~J.~Lipkin,
  Phys.\ Lett.\ B {\bf 172}, 242 (1986).
  doi:10.1016/0370-2693(86)90843-9.
	
\bibitem{zsgr} 
  S.~Zouzou, B.~Silvestre-Brac, C.~Gignoux and J.~M.~Richard,
  Z.\ Phys.\ C {\bf 30}, 457 (1986).
  doi:10.1007/BF01557611.
	
\bibitem{vvb} 
  J.~Vijande, A.~Valcarce and N.~Barnea,
  Phys.\ Rev.\ D {\bf 79}, 074010 (2009)
  doi:10.1103/PhysRevD.79.074010
  [arXiv:0903.2949 [hep-ph]].
	


\bibitem{lhcb} 
  R.~Aaij {\it et al.} [LHCb Collaboration],
  arXiv:1707.01621 [hep-ex].


\bibitem{Francis:2016hui} 
  A.~Francis, R.~J.~Hudspith, R.~Lewis and K.~Maltman, 
  Phys.\ Rev.\ Lett.\  {\bf 118}, 142001 (2017) 
  doi:10.1103/PhysRevLett.118.142001 
  [arXiv:1607.05214 [hep-lat]]. 


\bibitem{Bicudo:2016ooe} 
  P.~Bicudo, J.~Scheunert and M.~Wagner, 
  Phys.\ Rev.\ D {\bf 95}, 034502 (2017) 
  doi:10.1103/PhysRevD.95.034502 
  [arXiv:1612.02758 [hep-lat]]. 
	
\bibitem{kr} 
  M.~Karliner and J.~L.~Rosner,
  arXiv:1707.07666 [hep-ph].

\bibitem{eq} 
  E.~J.~Eichten and C.~Quigg,
  arXiv:1707.09575 [hep-ph].	
	
\bibitem{tHooft} 
  G.~'t Hooft,
  Nucl.\ Phys.\ B {\bf 72}, 461 (1974).
  doi:10.1016/0550-3213(74)90154-0
	
\bibitem{bln} 
  A.~V.~Berezhnoy, A.~V.~Luchinsky and A.~A.~Novoselov,
  Phys.\ Rev.\ D {\bf 86}, 034004 (2012)
  doi:10.1103/PhysRevD.86.034004
  [arXiv:1111.1867 [hep-ph]].
	
\bibitem{bkcv} 
  N.~Brambilla, G.~Krein, J.~Tarr\'us Castell\`a and A.~Vairo,
  Phys.\ Rev.\ D {\bf 93}, no. 5, 054002 (2016)
  doi:10.1103/PhysRevD.93.054002
  [arXiv:1510.05895 [hep-ph]].
	
\bibitem{knr} 
  M.~Karliner, S.~Nussinov and J.~L.~Rosner,
  Phys.\ Rev.\ D {\bf 95}, no. 3, 034011 (2017)
  doi:10.1103/PhysRevD.95.034011
  [arXiv:1611.00348 [hep-ph]].
	


\bibitem{Puchalski:2008jj} 
  M.~Puchalski and A.~Czarnecki,
  Phys.\ Rev.\ Lett.\  {\bf 101}, 183001 (2008)
  doi:10.1103/PhysRevLett.101.183001
  [arXiv:0810.0013 [hep-ph]].

	
\end{thebibliography}
\end{document}